\documentclass{PoS}

\usepackage{cleveref}

\graphicspath{{./plots/}}

\title{Inert Doublet Model Signatures at Future
           e$^+$e$^-$ Colliders}

\ShortTitle{Inert Doublet Model Signatures at Future 
           e$^+$e$^-$ Colliders}

\author{
  \speaker{Aleksander Filip \.Zarnecki},
        Jan Kalinowski, Jan Klamka, Pawel Sopicki\\
        Faculty of Physics, University of Warsaw\\
        E-mail: \email{filip.zarnecki@fuw.edu.pl},
        \email{jan.kalinowski@fuw.edu.pl},
        \email{j.klamka@student.uw.edu.pl},
        \email{pawel.sopicki@fuw.edu.pl}}

\author{Wojciech Kotlarski\\
        Institut f\"ur Kern- und Teilchenphysik, TU Dresden\\
        E-mail: \email{wojciech.kotlarski@tu-dresden.de}}

\author{Tania Robens\\
        Theoretical Physics Division, Rudjer Boskovic Institute, Zagreb\\
        E-mail: \email{trobens@irb.hr}}

\author{Dorota Sokolowska\\
        International Institute of Physics, Universidade Federal do Rio
  Grande do Norte,  Brasil\\
        E-mail: \email{dsokolowska@iip.ufrn.br}}

\abstract{
The Inert Doublet Model is one of the simplest extensions of the
Standard Model, providing a dark matter candidate. It is a two
Higgs doublet model with a discrete $Z_2$ symmetry, that prevents the
scalars of the second doublet (inert scalars) from coupling to the
Standard Model 
fermions and makes the lightest of them stable. We study a large
number of Inert Doublet Model scenarios, which are consistent with current
constraints on direct detection, including the most recent bounds
from the XENON1T experiment and relic density of dark matter, as well as
collider and low-energy limits. We use a set of benchmark
points with different kinematic features, that promise detectable
signals at future $e^+e^-$ colliders. Two inert scalar pair-production
processes are considered, $e^+e^- \to A~H $ and $e^+e^- \to H^+H^-$,
followed by decays of $H^\pm$ and $A$ into the final states 
which include the lightest and stable neutral scalar dark matter
candidate $H$. Significance of the expected observations is studied
for different benchmark models and different running scenarios, for
centre-of-mass energies up to 3\,TeV. Numerical results are presented
for the signal signatures  with two muons or an electron and a muon
in the final state, while the qualitative conclusions can also be
drawn for the semi-leptonic signatures. 
}

\FullConference{ALPS 2019 An Alpine LHC Physics Summit \\
		April 22 - 27, 2019\\
		Obergurgl, Austria}

\begin{document}

\section{Inert Doublet Model benchmark points}

\begin{figure}[b]
\includegraphics[width=0.49\textwidth]{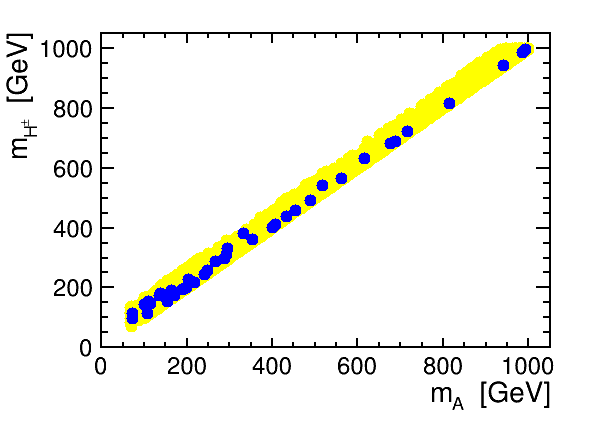}
\includegraphics[width=0.49\textwidth]{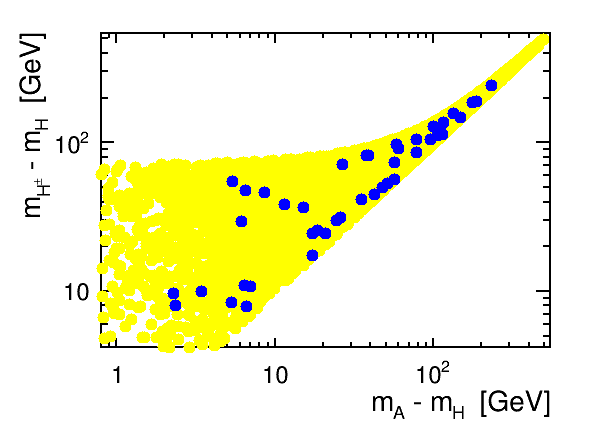}
\caption{ Distribution of benchmark candidate points (yellow) in the
  (m$_{A}$;m$_{H^\pm}$) plane (left) and in the
  (m$_{A} -\,$m$_{H}$;m$_{H^\pm} -\,$m$_{H}$) plane (right),
after all constraints are taken into account, as well as selected
benchmark points (blue) in the same planes~\cite{Kalinowski:2018ylg}.
}\label{fig:mass}
\end{figure}

One of the simplest extensions of the Standard Model (SM) which can
provide a dark matter (DM) candidate is the Inert Doublet Model
(IDM)~\cite{Deshpande:1977rw,Cao:2007rm,Barbieri:2006dq}.
In this model, the scalar sector is extended by a so-called inert or
dark doublet $\Phi_D$ (the only field odd under $Z_2$ symmetry) in
addition to the SM Higgs doublet $\Phi_S$. This results in five
physical states after electroweak symmetry breaking: the SM Higgs
boson $h$ and four dark scalars:  two neutral, $H$ and $A$, and two
charged,  $H^\pm$. 
Two sets of benchmark points in agreement with all
theoretical and experimental constraints were proposed
in~\cite{Kalinowski:2018ylg}, covering different possible signatures
at $e^+e^-$ colliders, with masses of IDM particles extending up to
1 TeV. 
Distributions of the scalar masses for the IDM benchmark scenarios
considered in~\cite{Kalinowski:2018ylg} are shown in Fig.~\ref{fig:mass}.
For the considered benchmark scenarios $H$ is the lightest, stable neutral
scalar and may serve as a good DM candidate.

\section{Analysis strategy}

\begin{figure}[tb]
\centerline{\includegraphics[width=0.75\textwidth]{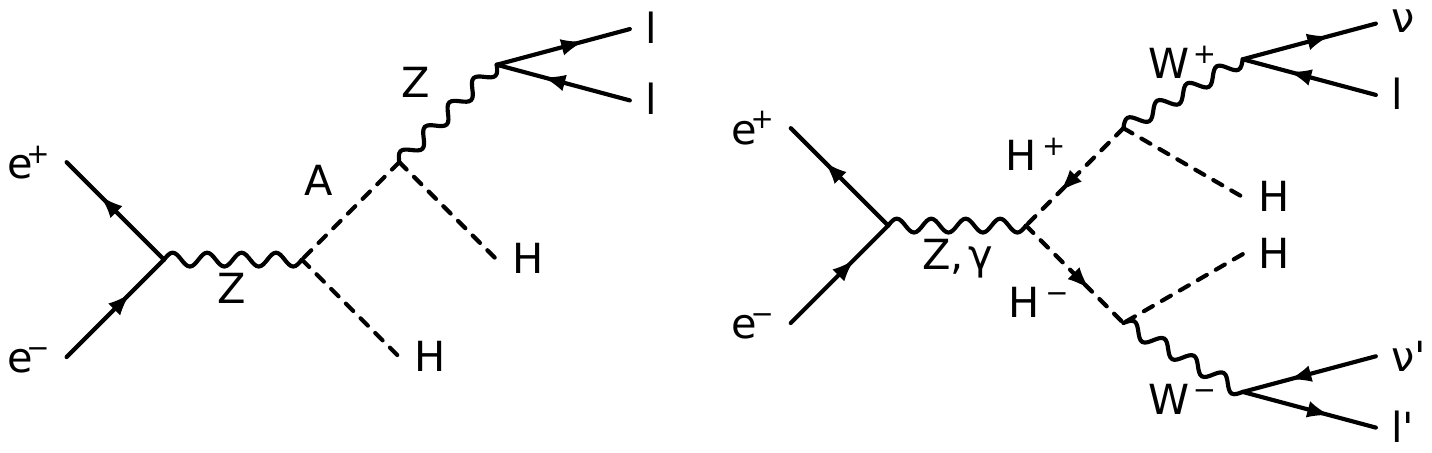}}
\caption{Signal Feynman diagrams for the considered production and
  decay process for:
{\sl (left)} neutral scalar production, $e^+e^- \to H A \to H H l l$,
and
{\sl (right)} charged scalar production, $e^+e^- \to H^+ H^- \to H H l l' \nu \nu'$.
}\label{fig:diag}
\end{figure}

Prospects for the discovery of IDM scalars at CLIC were described in
detail in~\cite{Kalinowski:2018kdn}. In this contribution we summarize
these results and extend them to ILC running at 250\,GeV and 500\,GeV.
The following tree-level production processes of inert scalars
at $e^+ e^-$ collisions are considered: $ e^+e^- \to A~H$
and $e^+e^-\to H^+H^-$.
In the scenarios considered in this paper the produced dark scalar $A$
decays to a (real or virtual) $Z$ boson and the (lighter)
neutral scalar $H$, $A \rightarrow Z^{(\star)}H$, while the produced
charged boson $H^\pm$ decays predominantly to a (real or virtual) $W^\pm$ boson
and the neutral scalar $H$, $H^+ \rightarrow {W^\pm}^{(\star)}H$,
{where the DM candidate $H$ escapes detection}.
Since isolated leptons (electrons and muons) can be identified and
reconstructed with very high efficiency and purity, we concentrate on
$Z$ and $W^\pm$ leptonic decays, leading to a signature of leptons and
missing transverse energy.
In the presented study, we considered the $\mu^+\mu^-$ final state as a
signature of the neutral scalar pair-production, while
the different flavour lepton pairs, $\mu^+ e^-$ and $e^+ \mu^-$, were
considered as a signature for production of charged inert scalars, see
Fig.~\ref{fig:diag}. 

\begin{figure}[tb]
\includegraphics[width=0.49\textwidth]{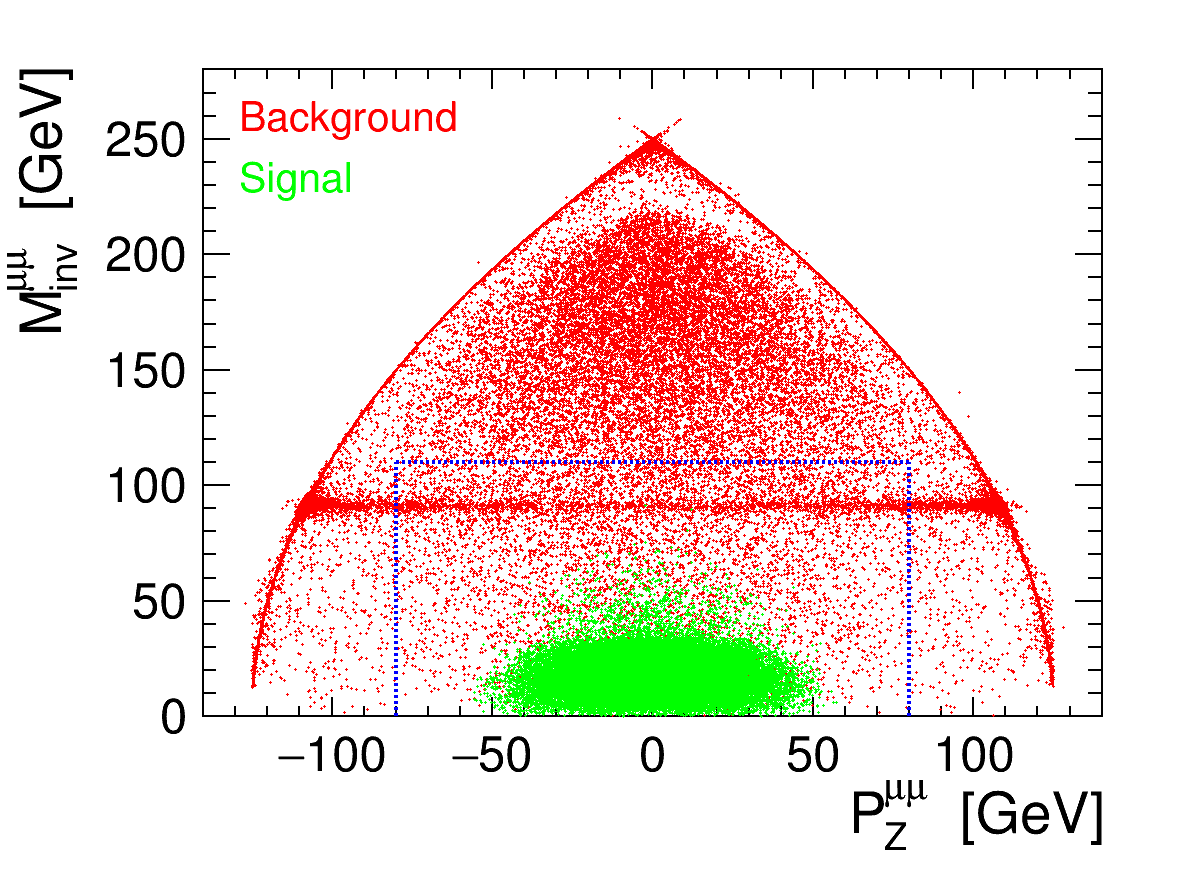}
\includegraphics[width=0.49\textwidth]{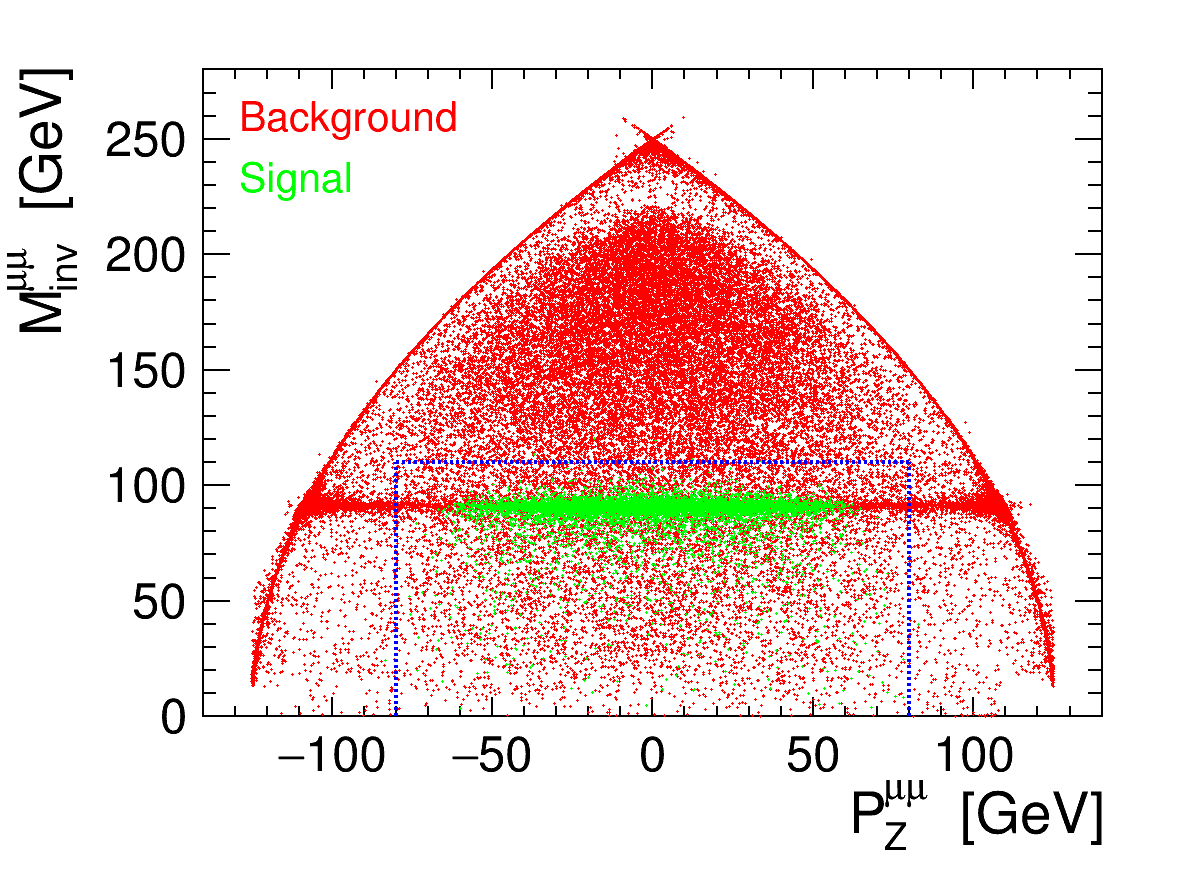}
\caption{
  Distribution of the lepton pair invariant mass, M$_{\mu\mu}$,
  as a function of the lepton pair longitudinal momentum,
  P$_\textrm{z}^{\mu\mu}$, for IDM signal (green points) and Standard Model
  background (red points). Signal events were simulated for BP1
  scenario (left) and BP9 scenario (right), for centre-of-mass energy
  of 250\,GeV. The blue box indicates the cut used to remove the
  dominant background from $e^+e^- \to \mu^+\mu^- (\gamma)$ process.
}\label{fig:presel}
\end{figure}

\begin{figure}[tb]
\includegraphics[width=0.49\textwidth]{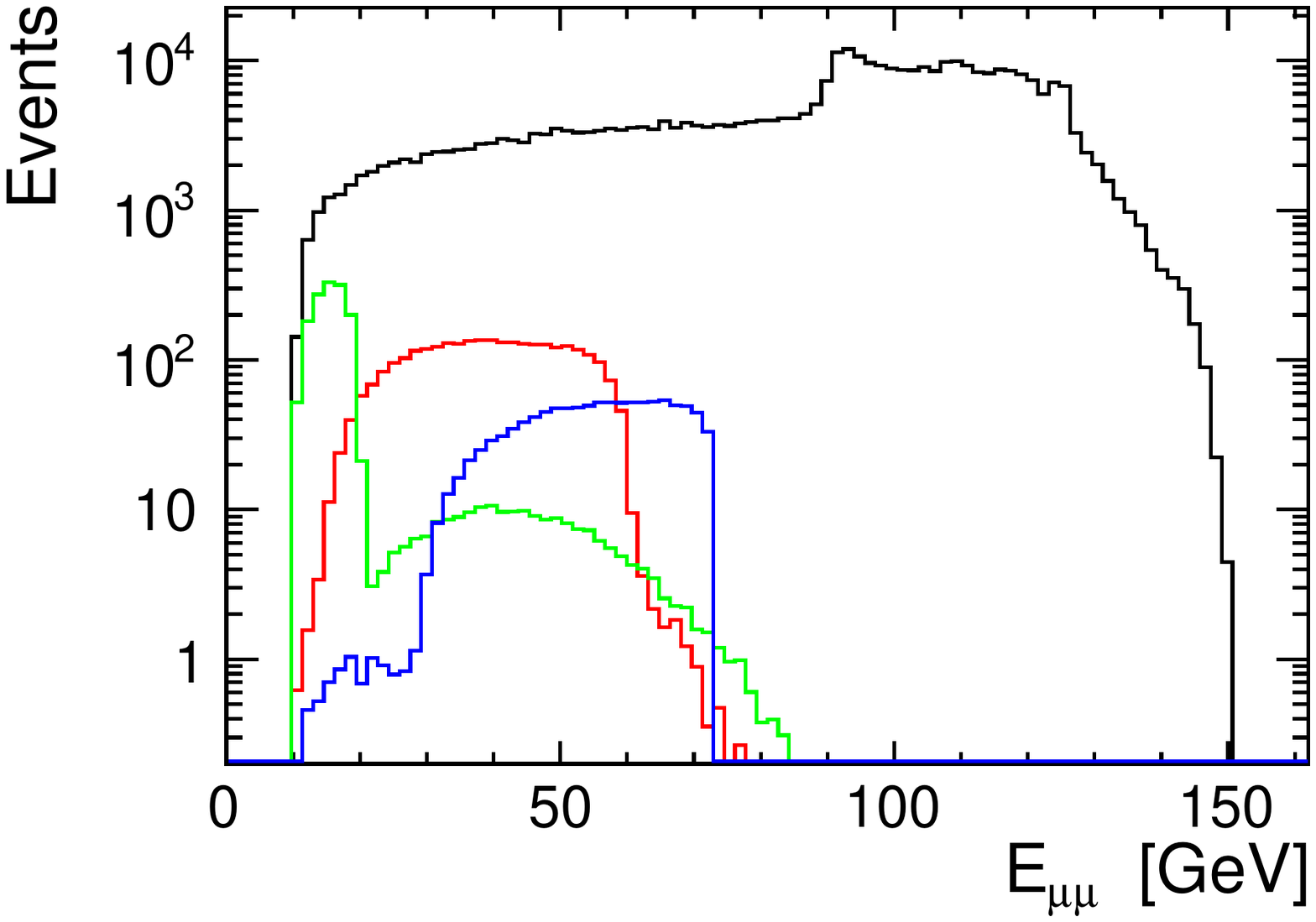}
\includegraphics[width=0.49\textwidth]{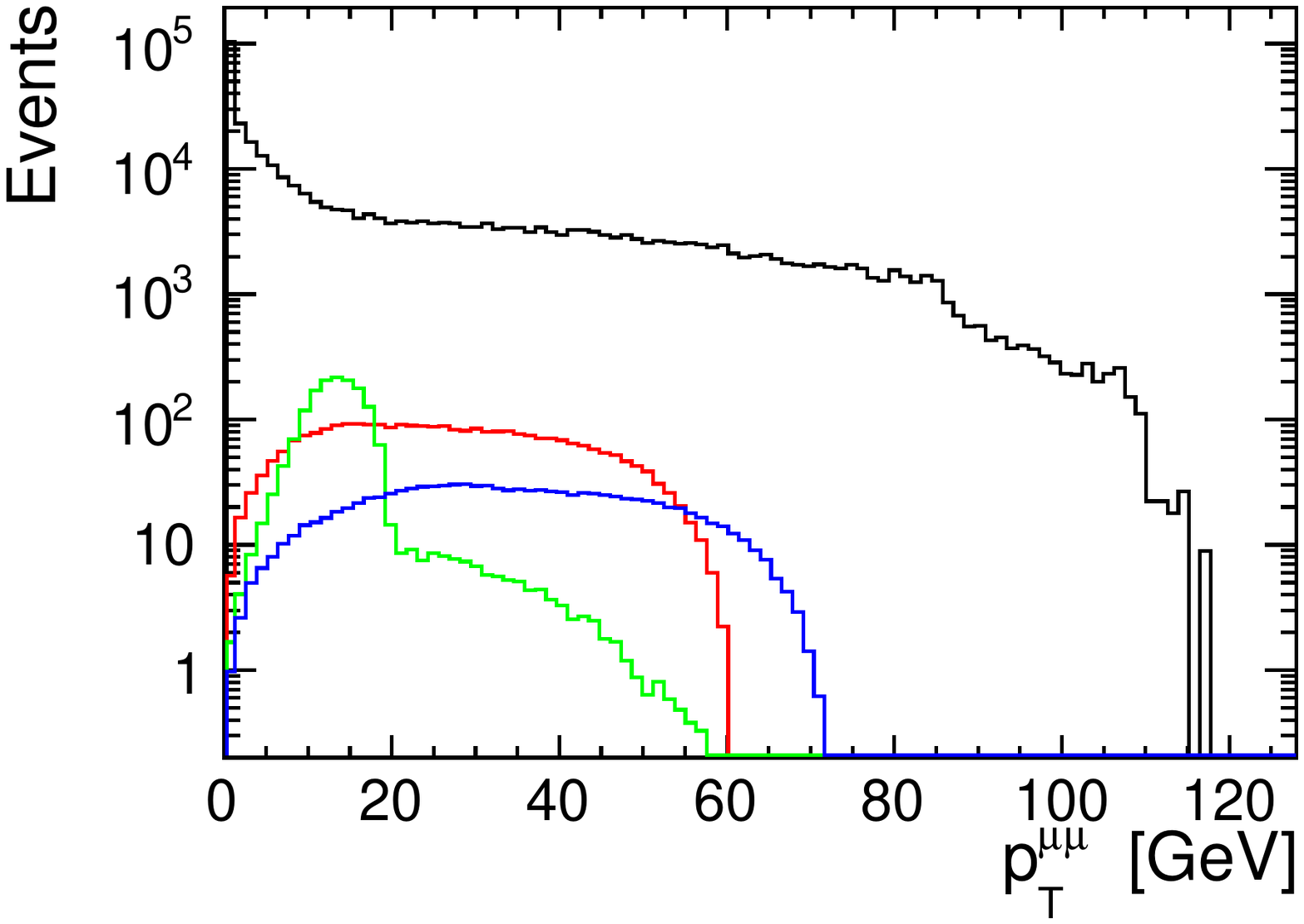}
\caption{
  Distributions of the kinematic variables describing the
  leptonic final state in $AH$ analysis: lepton pair
energy, E$_{\mu\mu}$ and total transverse momentum, p$^{\mu\mu}_\textrm{T}$.
Expected distributions for representative benchmarks BP1 (red
histogram), BP2 (green) and BP7 (blue) are compared with expected
background (black histogram) simulated for 1\,ab$^{-1}$ collected at 250\,GeV.
}\label{fig:dist}
\end{figure}

\begin{figure}[tb]
\includegraphics[width=0.49\textwidth]{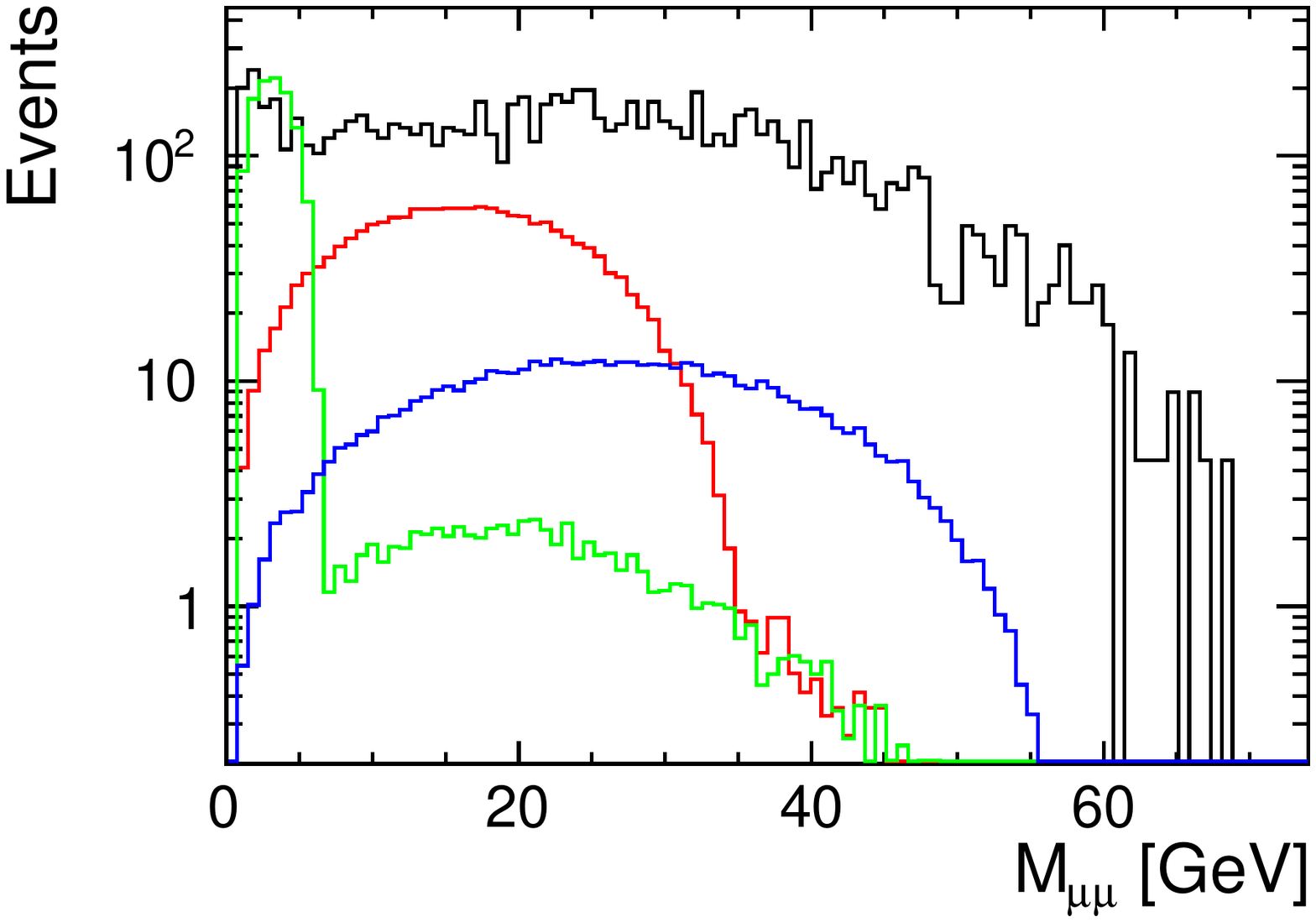}
\includegraphics[width=0.49\textwidth]{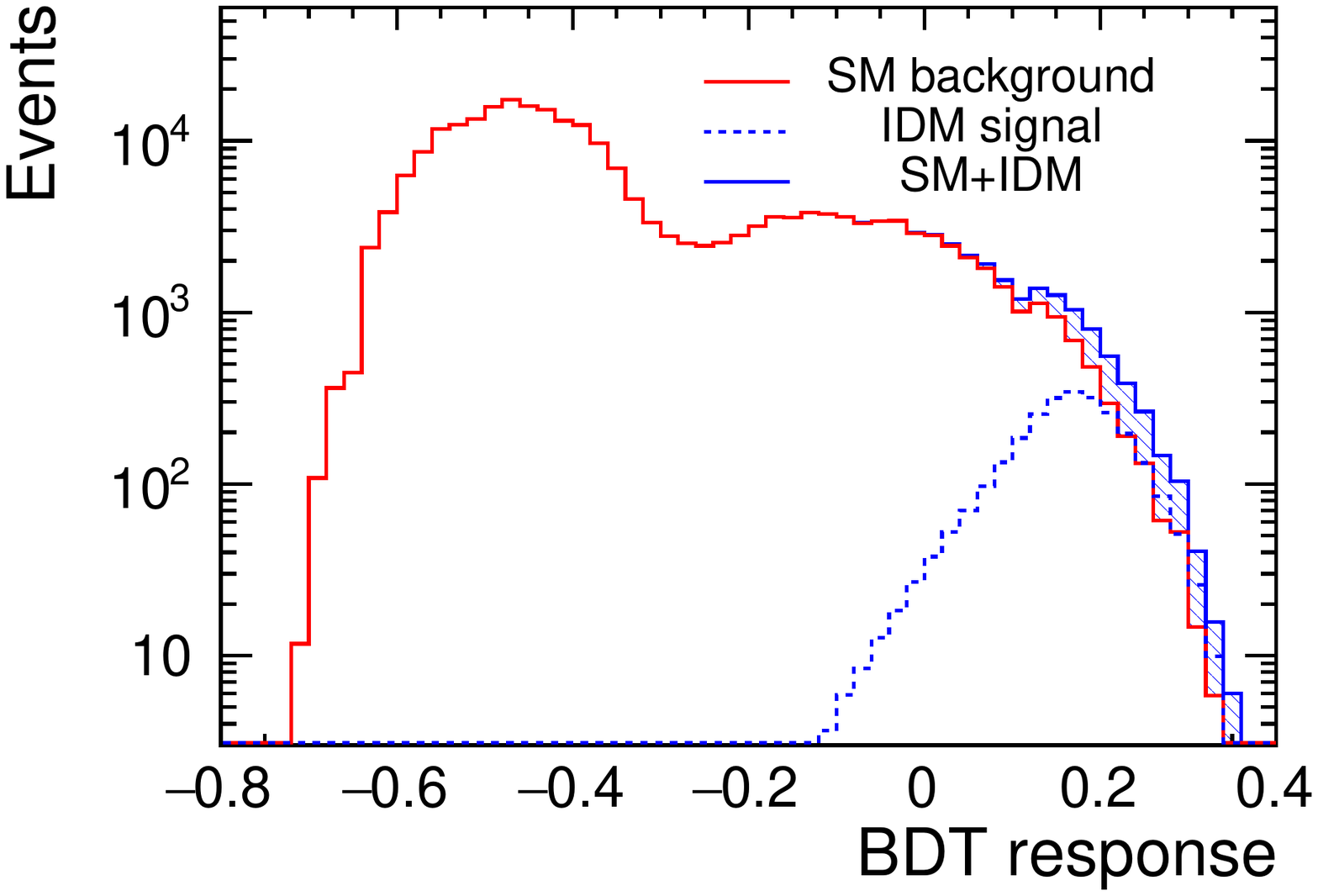}
\caption{
  Left: distribution of the lepton pair invariant mass, M$_{\mu\mu}$,
  for BP1 (red histogram), BP2 (green) and BP7 (blue) signal
  scenarios, compared with the expected Standard Model background (black
  histogram), after event  selection cuts (see text for details).
  Right: response distributions of the BDT classifiers used for the
  selection of $AH$ production events, for BP1.
  Samples are normalised to 1\,ab$^{-1}$ collected at 250\,GeV.
}\label{fig:bdt}
\end{figure}

Signal and background samples were generated with WHizard
2.2.8~\cite{Kilian:2007gr}. Generator level cuts reflecting detector
acceptance for leptons and ISR photons were applied.
For the neutral inert scalar pair production, $e^+ e^- \to AH$,
the invariant mass of the lepton pair from (virtual) $Z$ decay depends on the
mass splitting between $A$ and $H$ and is relatively small,
$M_{\mu\mu} \le M_{Z}$. 
We apply pre-selection cuts on the invariant mass  and the
longitudinal boost of the lepton pair to suppress the dominant
background process $e^+ e^- \to \mu^+ \mu^- (\gamma)$, see
Fig.~\ref{fig:presel}. 
Distributions of selected kinematic variables describing the
leptonic final state in $AH$ analysis, after the pre-selection cuts,
are presented in Fig.~\ref{fig:dist}.
Presented in Fig.~\ref{fig:bdt}\,(left) is the lepton pair invariant mass
distribution after pre-selection cuts and additional selection based
on lepton pair energy, transverse momentum, production angle (polar
angle of the Z boson) and the difference of the lepton azimuthal angles.
Already with this simplest, cut-based approach, the IDM signal would
result in the visible excess in the invariant mass distribution for
the number of benchmark scenarios.
For the final selection of signal-like events, a multivariate analysis
is performed using a Boosted Decision Tree (BDT) classifier
\cite{Hocker:2007ht} with 8 input variables, both for $HA$ and $H^+
H^-$ analysis (no pre-selection cuts are applied for $H^+H^-$).
The BDT is trained using all accessible (at given energy) benchmark
points in given category ($\mu^+\mu^-$ or $e^\pm \mu^\mp$ signature;
virtual or real $W$/$Z$). 
Response distributions of the BDT classifier used for the
selection of $AH$ production events for the benchmark scenario BP1 are
presented in Fig.~\ref{fig:bdt}\,(right).

\section{Results}

\begin{figure}[tb]
\centerline{\includegraphics[width=0.8\textwidth]{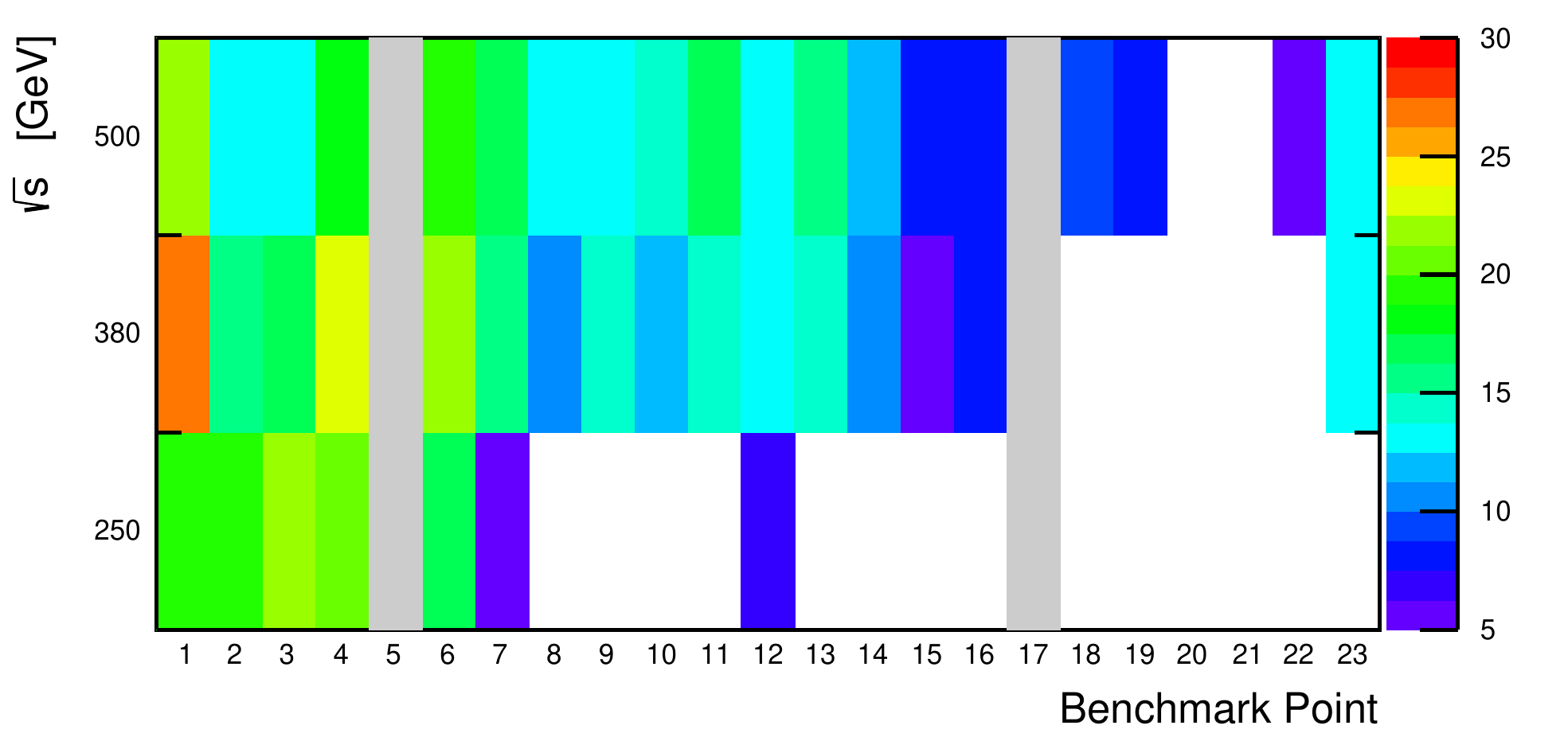}}
\caption{Significance of the deviations from the Standard Model
  predictions, expected for 1\,ab$^{-1}$
  of data collected at centre-of-mass energy of 250\,GeV, 380\,GeV and
  500\,GeV, for events with two muons in the final state, for all considered
  low mass benchmark scenarios. Only significance above 5$\sigma$ is shown.
}\label{fig:ahsig}
\end{figure}

\begin{figure}[tb]
\centerline{\includegraphics[width=0.8\textwidth]{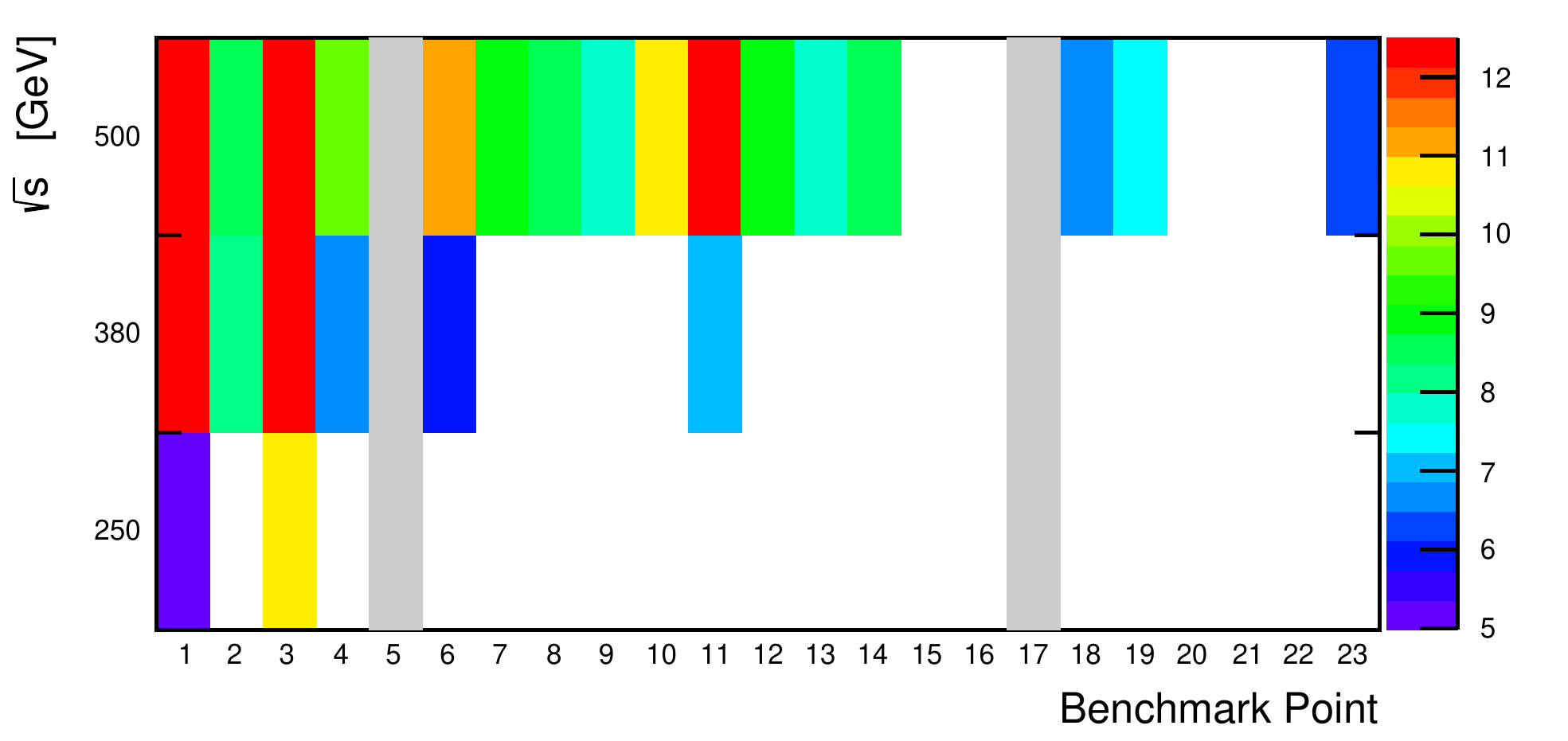}}
\caption{As in Fig.~\ref{fig:ahsig} but for events  with an electron
  and a muon in the final state.
}\label{fig:hphmsig}
\end{figure}

\begin{figure}[tb]
\hspace{0.03\textwidth}
  \includegraphics[width=0.47\textwidth]{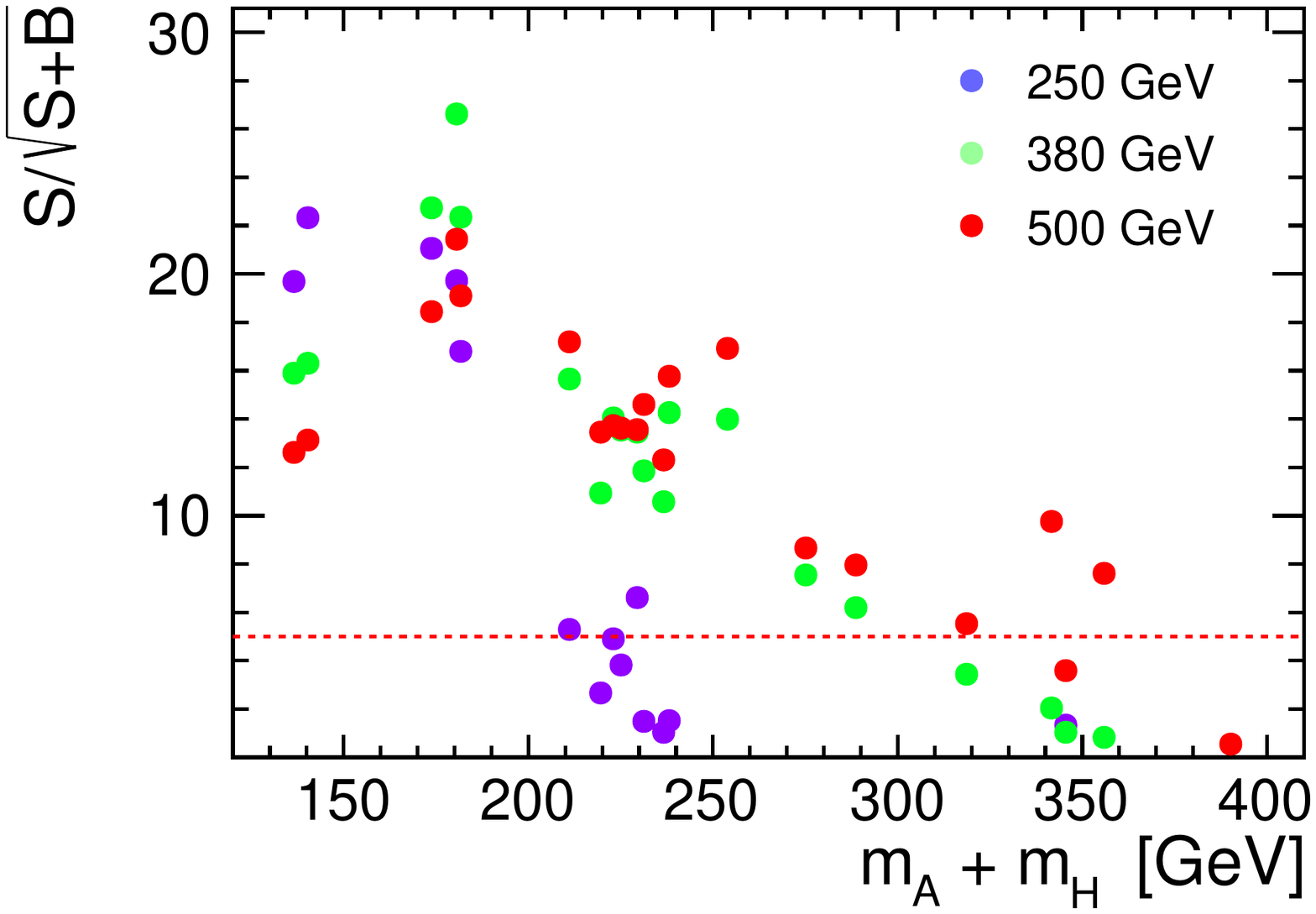}
  \includegraphics[width=0.47\textwidth]{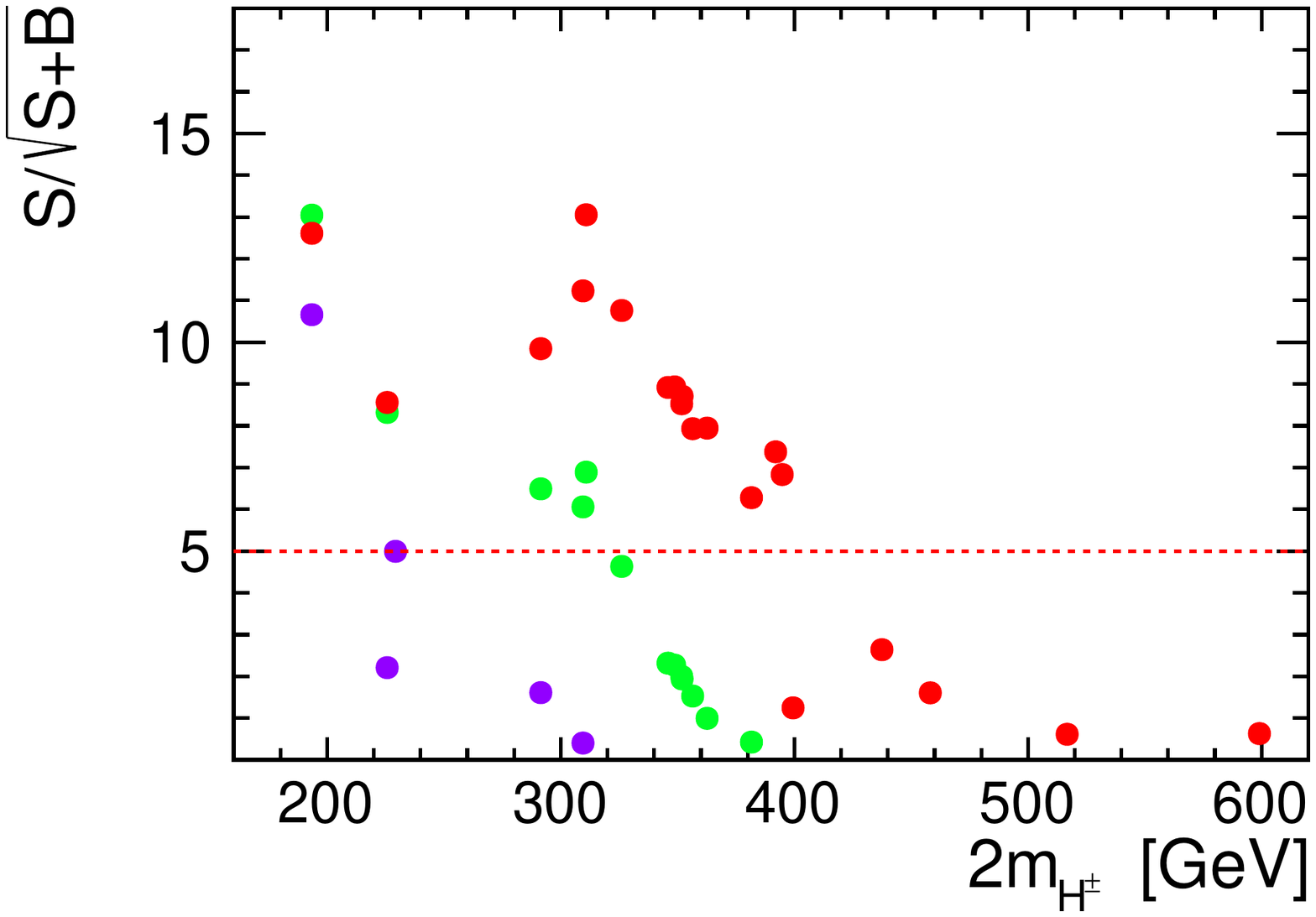}
\caption{Significance of the deviations from the Standard Model
  predictions, expected for 1\,ab$^{-1}$ of data collected at
  centre-of-mass energy of 250\,GeV, 380\,GeV and   500\,GeV, for:
  {\sl (left)} events with two muons in the final state ($\mu^+\mu^-$)
  as a function of the sum of neutral inert scalar masses and {\sl
    (right)} events with an electron and a muon in the final state
  ($e^+\mu^-$ or $e^-\mu^+$) as a function of twice the charged scalar
  mass.  
}\label{fig:lesig}
\end{figure}

Expected significance of the deviations from the Standard Model 
predictions, assuming  1\,ab$^{-1}$ of data collected at
centre-of-mass energy of 250\,GeV, 380\,GeV and 500\,GeV, for $AH$ and
$H^+H^-$ signatures, are shown in Figs.~\ref{fig:ahsig} and
\ref{fig:hphmsig}, respectively.
Only scenarios resulting in the significances above 5$\sigma$ are
shown\footnote{We omit BP5 and BP17, which are excluded by the updated
  Xenon1T limits.}. 
We found that for scenarios accessible at a certain energy, high
significances can be expected at future $e^+e^-$ colliders.
They are mainly related to the inert scalar production cross
sections.
We display the dependence of the expected significances on the inert
scalar masses is Fig.~\ref{fig:lesig}.
With  1\,ab$^{-1}$ of integrated luminosity collected at 250\,GeV,
380\,GeV and   500\,GeV, the expected discovery reach of $e^+e^-$
colliders extends up to neutral scalar mass sum of 220\,GeV, 300\,GeV
and  330\,GeV, respectively, and for charged scalar pair-production up
to charged scalar masses of 110\,GeV, 160\,GeV and  200\,GeV.
\begin{figure}[tb]
 \hspace{0.03\textwidth}
  \includegraphics[width=0.47\textwidth]{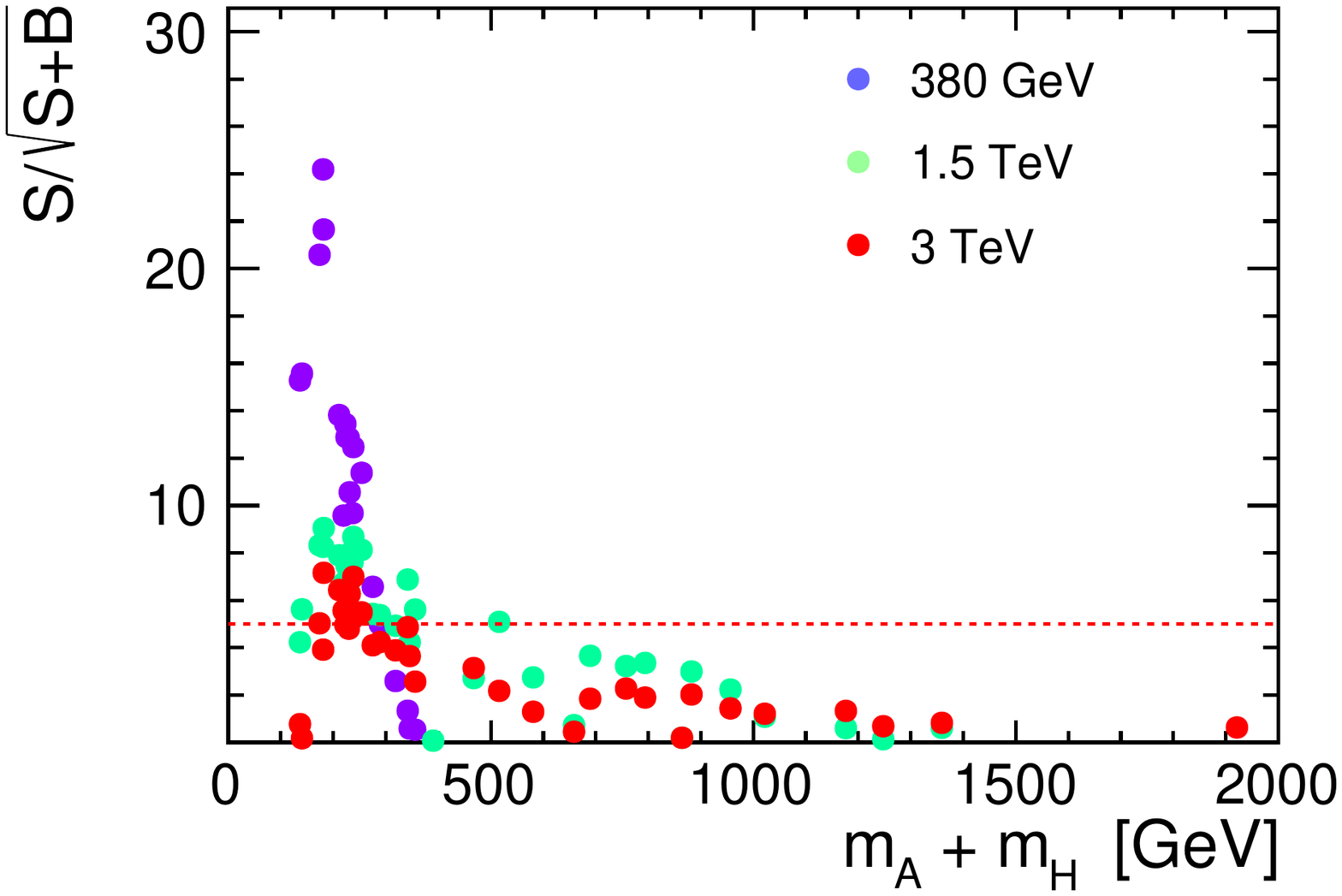}
  \includegraphics[width=0.47\textwidth]{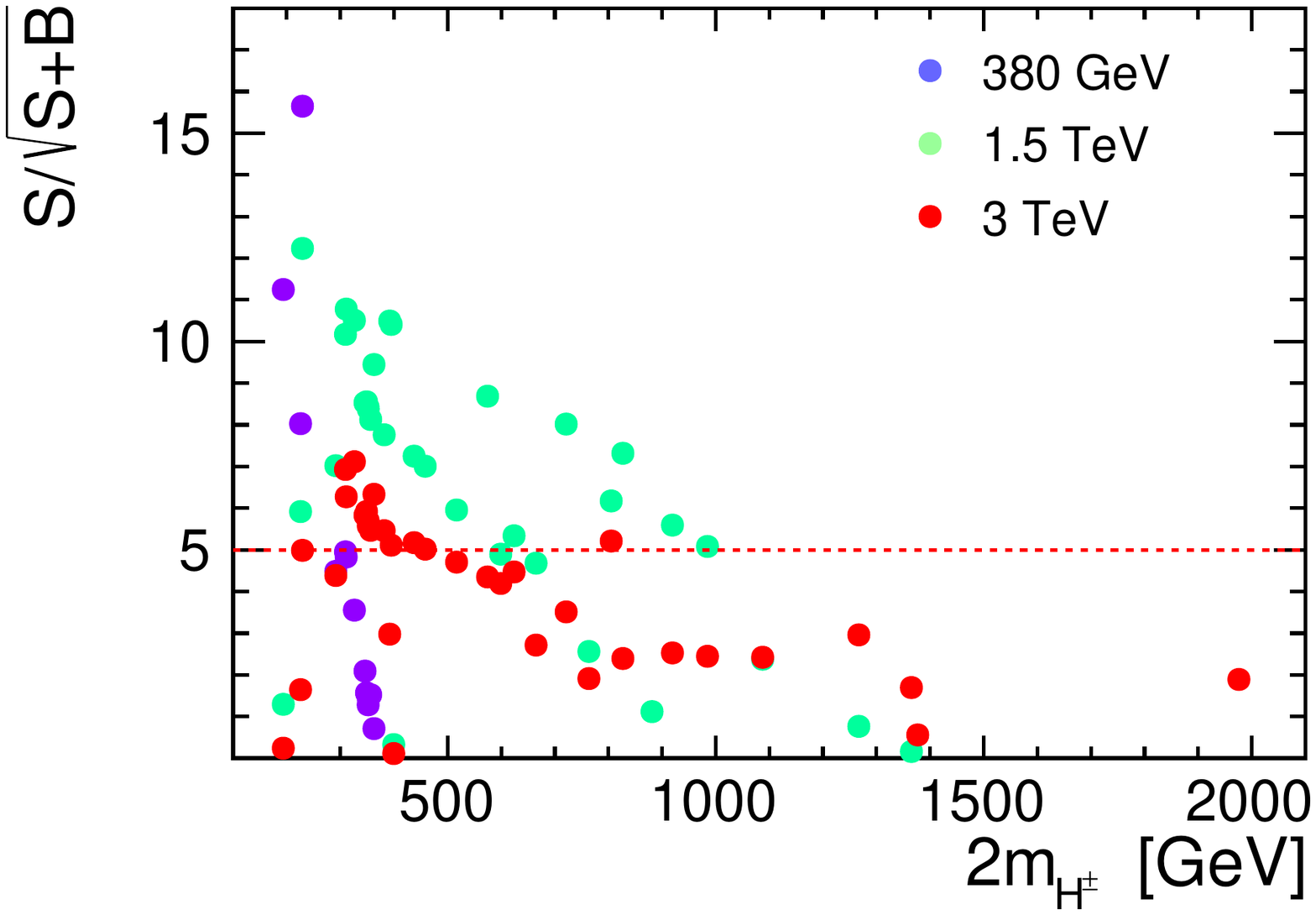}
  \caption{As in Fig.~\ref{fig:lesig} but for expected CLIC running
    scenario: 1\,ab$^{-1}$ of data collected at 380\,GeV,
    2.5\,ab$^{-1}$ at 1.5\,TeV and 5\,ab$^{-1}$ at 3\,TeV.
  }\label{fig:hesig}
\end{figure}
For CLIC running at 1.5\,TeV, only a moderate increase in discovery
reach is expected, even with 2.5\,ab$^{-1}$ of data, see
Fig.~\ref{fig:hesig}.
The neutral scalar pair-production can be discovered in the leptonic
channel for $m_A + m_H < 450$\,GeV
and the charged scalar production for $m_{H^\pm} < 500$\,GeV.
Marginal improvement is expected when running at 3 TeV.
Low significance expected at high CLIC energies is related to
the decrease of the signal cross sections with energy.
However, we found that significant improvement of the discovery reach
for high mass scenarios could be achieved using the semi-leptonic final state,
as the signal cross section increases by an order of magnitude.
The corresponding study is in progress and the preliminary results are
very promising.

\subsection*{Acknowledgements}

This contribution was supported by the National Science Centre, Poland, the
OPUS project under contract UMO-2017/25/B/ST2/00496 (2018-2021) and
 the HARMONIA project under contract UMO-2015/18/M/ST2/00518
 (2016-2019).

\end{document}